%% file: main.tex
\documentclass[
]{ceurart}

\usepackage{listings}
\input{macros.tex}

\begin{document}

\copyrightyear{2022}
\copyrightclause{Copyright for this paper by its authors.
  Use permitted under Creative Commons License Attribution 4.0
  International (CC BY 4.0).}

\conference{Green-Aware AI 2025 @ ECAI 2025}

\title{Choosing to Be Green: Advancing Green AI via Dynamic Model Selection} %

\author[1]{Emilio Cruciani}[%
orcid=0000-0002-4744-5635,
email=emilio.cruciani@unier.it,
url=https://sites.google.com/view/emiliocruciani,
]
\address[1]{European University of Rome, Italy}

\author[2]{Roberto Verdecchia}[%
orcid=0000-0001-9206-6637,
email=roberto.verdecchia@unifi.it,
url=https://robertoverdecchia.github.io,
]
\address[2]{University of Florence, Italy}

\begin{abstract}
Artificial Intelligence is increasingly pervasive across domains, with ever more complex models delivering impressive predictive performance. 
This fast technological advancement however comes at a concerning environmental cost, with state-of-the-art models---particularly deep neural networks and large language models---requiring substantial computational resources and energy.
In this work, we present the intuition of \textit{Green AI dynamic model selection}, an approach based on dynamic model selection that aims at reducing the environmental footprint of AI by selecting the most sustainable model while minimizing potential accuracy loss. 
Specifically, our approach takes into account the inference task, the environmental sustainability of available models, and accuracy requirements to dynamically choose the most suitable model. Our approach presents two different methods, namely \textit{Green AI dynamic model cascading} and \textit{Green AI dynamic model routing}. We demonstrate the effectiveness of our approach \textit{via} a proof of concept empirical example based on a real-world dataset. 
Our results show that \textit{Green AI dynamic model selection} can achieve substantial energy savings (up to $\approx$25\%) while substantially retaining the accuracy of the most energy greedy solution (up to $\approx$95\%). 
As conclusion, our preliminary findings highlight the potential that hybrid, adaptive model selection strategies withhold to mitigate the energy demands of modern AI systems without significantly compromising accuracy requirements.
\end{abstract}


\maketitle

\input{trunk/intro}

\input{trunk/related}
\input{trunk/approach}
\input{trunk/example}

\bibliography{biblio}

\end{document}

%% file: macros.tex
\usepackage{algorithm}
\usepackage[noend]{algpseudocode}

\usepackage{siunitx}
\usepackage{amsmath}
\usepackage{xspace}

%% file: trunk/intro.tex
\section{Introduction}\label{sec:intro}
The popularization of AI models, ranging from simple classifiers to complex large language models, has taken the world by storm. With the widespread and evergrowing adoption of AI and all the benefits it implied, the environmental resources needed to power such models is also surging, and this trend is no longer negligible~\cite{wu2022sustainable}. To contrast the invisible impact that AI is having on the limited resources of our planet, the field of \textit{Green AI}~\cite{schwartz2020green} rapidly developed, and has seen a considerable growth in the most recent years~\cite{verdecchia2023systematic}. By quoting the words of Schwartz~et~al.~\cite{schwartz2020green}, Green AI is a field of AI research that yields novel results while considering its computational cost and encouraging the reduction of resources spent. Under the research field of Green AI fall a plethora of heterogeneous solutions, ranging from \textit{ad hoc} hyperparameter tuning to trade-offs between model accuracy and energy consumption, energy-aware model deployment strategies, data-centric techniques~\cite{verdecchia2023systematic}, and software engineering approaches~\cite{cruz2025greening}. Despite the wide array of Green AI solutions that have been conceived to date, Green AI approaches based on selecting different models by factoring in their energy consumption results to date to be an uncharted territory. In this work, we explore the potential that dynamic model selection based on the task at hand, model validation accuracy, and energy efficiency can have on AI environmental sustainability. More specifically, we present the very idea of \textit{Green AI dynamic model selection} by presenting two methods that lend their core intuition from the related literature on dynamic model selection, namely \textit{model cascading} and \textit{model routing}~\cite{viola2001cascade,jacobs1991moe}. Intuitively the first method we present, namely \textit{Green AI dynamic model cascading}, subsequently invokes different models from less to more energy greedy till a sufficient level of prediction confidence is achieved. The second method instead, referred to as \textit{Green AI dynamic model routing}, selects exclusively the most efficient model to be used by considering the predicted model accuracy for the input task and the energy efficiency of the model. In addition to a formal presentation of both energy-aware dynamic model cascading and routing methods, we also document the results of an empirical proof of concept evaluation that we execute to showcase the viability of our intuition. The empirical proof of concept, which should be by no means be interpreted as exhaustive or conclusive, is based on an exemplary classification task relying on the widely utilized scikit-learn and keras Python libraries, two AI models of different complexity, and an \textit{ad hoc} implementation of the energy-aware dynamic model cascading and routing methods. The preliminary results we collect point to the potential that dynamic model selection methods have to achieve Green AI. As a complementary portion of our contribution, we also delve into reflecting on the various nuances, potential challenges, and benefits that may arise when building further onto the \textit{Green AI dynamic model selection} approach.

%% file: trunk/related.tex
\section{Related Work}\label{sec:related}
The field of Green AI has experienced a swift growth in popularization in the past few years and is increasingly becoming an established discipline~\cite{verdecchia2023systematic}. The rise of interest in the topic could have stemmed from diverse effort of the research community to quantify the environmental impact of AI, ranging from generic high level figures of CO$_2$ emissions~\cite{lacoste2019quantifying} to fine-grained measurements of specific models, e.g., deep learning ones~\cite{strubell2020energy}. The overall picture studies of this nature draw is consistent, the environmental impact of AI is an issue that needs to be addressed. Answering such call, numerous research endeavors focused on improving the energy efficiency and environmental sustainability of AI. The proposed solutions to achieve Green AI are heterogeneous and span a wide range of approaches. 

A family of Green AI techniques focuses on designing AI models by factoring in their energy consumption~\cite{rouhani2016delight}, e.g., by improving model execution times~\cite{garcia2021energy}, optimizing models for specific hardware components~\cite{rungsuptaweekoon2017evaluating},  compressing models~\cite{yang2018energy}, or seeking more energy efficient model implementations~\cite{11039308, georgiou2022green}. In contrast, another family of Green AI techniques focus instead on  the \textit{a posteriori} optimization of models \textit{via} hyperparameter fine-tuning~\cite{de2021hyperparameter, magno2017deepemote, barlaud2021learning, stamoulis2018designing, sponner2024adapting}.

As a green AI research area that might somewhat be more related to the topic considered in this research, a set of studies investigated how different model deployment strategies can impact their energy consumption. Contributions of this type consider solutions such as inference on the edge~\cite{gondi2021performance, yang2020sparse}, model deployment in virtualized cloud fog networks~\cite{yosuf2021energy}, and distributed machine learning~\cite{guler2021energy}.

Taking a different standpoint, other Green AI approaches consider instead exclusively the data the models are trained with, rather than the design of the algorithm themselves, a discipline referred to as \textit{Data Centric Green AI}~\cite{verdecchia2022data, salehi2023data, alswaitti2025training, kumar2024opportunities, jarvenpaa2024synthesis}.

All of the above mentioned areas of Green AI research result orthogonal to the topic considered in this study, as they focus exclusively on the optimization of one specific model. In contrast, in our work, we do not aim to improve the energy efficiency of a single model, but rather to select the most fitting one (or set thereof) by keeping AI energy efficiency in mind. To the best of our knowledge, this topic has to date only marginally be explored in the related literature.

The work of~\citet{nijkamp2024green} is potentially the one that is most closely related to the approaches presented in this contribution. Nijkamp et al. consider an ensemble learning context within the text processing domain, where a subset of pre-trained and trained models are selected for inference and results are merged \textit{a posteriori}. The selection of models can be executed either statically, where an optimal subset of models is chosen for an entire domain considered, or dynamically, i.e., an optimal subset is selected for every queried property within the domain. Differently from such approach, we do not focus on ensemble learning, and trigger the inference of multiple models only in the case of energy-aware model cascading (see also Section~\ref{sec:cascading}). 

In another work that considers ensemble learning, \citet{omar2024more} consider the impact that three different design decisions for ensemble learning, namely ensemble size, fusion methods, and partitioning methods, can have on energy consumption. As for the previous study, our contribution differs by not considering the context of ensemble learning, but rather dynamic model selection for energy efficiency. A related work by \citet{matathammal2025edgemlbalancer} presents EdgeMLBalancer, an approach that balances resource utilization \textit{via} dynamic model switching in the context of edge-devices. In contrast to such work, our dynamic model selection approach is not concerned with the allocation between different resource-constrained edge devices, is not specific to real-time object detection, and is not based on the MAPE-K Feedback Loop to conduct the selection of models (see also Section~\ref{sec:approach}). 

As mention of another branch of related work, at the core of this study lies a plethora of foundational research endeavors conducted in the realm of model selection~\cite{anderson2004model, merz1996dynamical, armstrong2006dynamic}, with particular emphasis on  approaches based on model cascading and model routing~\cite{viola2001cascade,jacobs1991moe}. Our contribution builds upon such literature, by borrowing the intuition of such approaches to embed environmental sustainability as part of the dynamic model selection process.

%% file: trunk/approach.tex
\section{Green AI Dynamic Model Selection Methods}\label{sec:approach}

\begin{figure}
    \centering
    \includegraphics[width=1\linewidth]{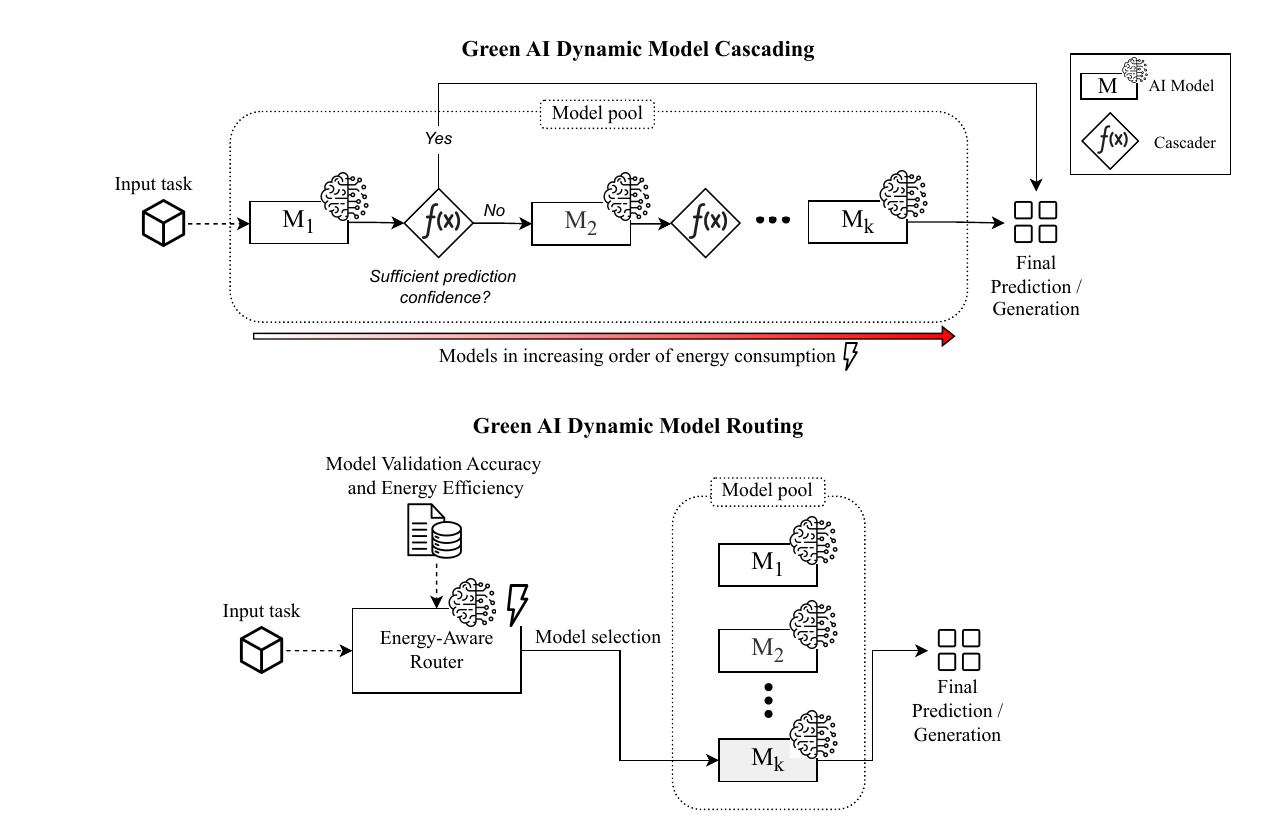}
    \caption{Overview of the dynamic model selection methods for energy efficient inference.}
    \label{fig:approaches}
\end{figure}

In this section we present two methods to build energy aware classification models \textit{via} cascading and routing. An overview of the proposed dynamic model selection methods for
energy efficiency are depicted in Figure~\ref{fig:approaches} and are further described below. 

Intuitively the first strategy, named \textit{Green AI dynamic model cascading}, is based on a cascading  methods where models at an increasing level of energy consumptions are invoked subsequently when required. The second method instead, named \textit{Green AI dynamic model routing}, is based on an upfront energy-aware router component that selects the best suited model based on the task at hand, the validation accuracy of models, and their energy efficiency.

As a note on terminology, in the following documentation both methods we present consider a labeled dataset $D = (\mathbf{x}_i, y_i)_{i=1\ldots,n}$ of $n$ data points with $\mathbf{x}_i \in X$ being the feature vector of data point $i$ and $y_i \in Y$ being its label.
Also, we let $M$ be a classification model, namely a function $M : X \to Y$ that given an input $\mathbf{x} \in X$ predicts its class as $\hat{y} = M(\mathbf{x}) \in Y$.

\subsection{Green AI Dynamic Model Cascading}
\label{sec:cascading}

For the cascading model $C : X \to Y$, we need: 
a sequence of $k \ge 2$ models $M_1, M_2, \ldots, M_k$, ordered by increasing energy consumption and typically increasing complexity and accuracy; 
a family of prediction confidence functions $\alpha_i(\mathbf{x}) : X \times Y \to [0,1]$, depending on each model $M_i$;
a parameter $\epsilon \in [0, 1]$ to control the confidence tolerance.

At inference time, for an input instance $\mathbf{x}$, the cascading mechanism proceeds as described in Algorithm~\ref{alg:cascading}.
In particular, the cascading model evaluates the first model $M_1$ and obtains both the predicted label $M_1(\mathbf{x})$ and its confidence score $\alpha_1(\mathbf{x})$.
If the confidence satisfies $\alpha_1(M_1(\mathbf{x})) \geq 1 - \epsilon$ for some $\epsilon \in (0,1)$, we accept the prediction and terminate.
Otherwise, we move to the next model $M_2$ and repeat the procedure.
This continues until either a model $M_i$ produces a sufficiently confident prediction or the first $k-1$ models are exhausted, in which case we use the last model $M_k$ as a fallback.

\begin{algorithm}
\caption{Energy-Aware Cascading Inference}
\begin{algorithmic}[1]
\Require{Instance $\mathbf{x}$}
\For{$i = 1$ to $k-1$}
    \If{$\alpha_i(\mathbf{x}) \geq 1 - \epsilon$}
        \State \Return $M_i(\mathbf{x})$
    \EndIf
\EndFor
\State \Return $M_k(\mathbf{x})$
\end{algorithmic}
\label{alg:cascading}
\end{algorithm}

\subsection{Green AI Dynamic Model Routing}

An alternative method to reduce inference cost is to directly learn a routing function that selects, for each input, the most appropriate model in terms of energy-accuracy tradeoff. 
In this case, we define a routing model \( R : X \to \{1, \ldots, k\} \), which maps each input instance to one of the \( k \) available models \( M_1, M_2, \ldots, M_k \), again ordered by increasing energy consumption.

At inference time, for a given input \( \mathbf{x} \in X \), the routing model selects an index \( i = O(\mathbf{x}) \) and returns the prediction \( M_i(\mathbf{x}) \), as described in Algorithm~\ref{alg:routing}.

\begin{algorithm}
\caption{Energy-Aware Routing Inference}
\begin{algorithmic}[1]
\Require{Instance $\mathbf{x}$}
\State $i \gets O(\mathbf{x})$
\State \Return $M_i(\mathbf{x})$
\end{algorithmic}
\label{alg:routing}
\end{algorithm}

The goal is to train a model \( O \) so that it selects the least energy-consuming model capable of producing a correct prediction.
To train the routing model, we assume access to a validation dataset \( D_{\text{val}} \subseteq D \), and we construct training labels for the routing task as follows: for each input \( \mathbf{x} \in D_{\text{val}} \), we identify the lowest-index model \( M_i \) that correctly classifies \( \mathbf{x} \) (i.e., \( M_i(\mathbf{x}) = y \)); if no such model exists, we select the lowest-index model regardless of accuracy to minimize energy cost. 
This label construction assumes that we can evaluate the correctness of each model on the validation set and that the energy consumption associated with each model is known.

Formally, we define an oracle routing function \( O^* : X \to \{1, \ldots, k\} \) that given \( \mathbf{x} \) returns:
\[
    O^*(\mathbf{x}) = 
    \begin{cases}
    \min \left\{ i \in \{1,\ldots,k\} \,:\, M_i(\mathbf{x}) = y \right\} & \text{if } \exists i \text{ s.t. } M_i(\mathbf{x}) = y,
    \\
    1 & \text{otherwise}.
    \end{cases}
\]
We then train the model \( O \) to approximate the oracle \( O^* \), using standard classification techniques. 

%% file: trunk/example.tex
\section{Empirical Proof of Concept}\label{sec:example}

In this section, we demonstrate the advantages of our approach through a concrete example, comparing the performance and energy consumption of our dynamic model selection classifiers with those of their basic components.
As described in Section~\ref{sec:approach}, we consider a cascading model $C$ and a routing model $R$.

We consider the standard multi-class classification task on
the scikit-learn \texttt{digits} dataset\footnote{\url{https://scikit-learn.org/1.5/auto_examples/datasets/plot_digits_last_image.html}}.
The dataset consists of 1797 8x8 grayscale images of hand-written digits (0-9).
We use 60\% of the dataset for training, a 20\% for validation, and 20\% for testing.
The validation set is only used by the routing model $R$ to training the oracle, while it is not used by the cascading model $C$ for a fair comparison.

For simplicity, in this proof of concept we use only two components for each of the two models:
a shallow (depth 5) decision tree $G$ (the simpler and greener model), 
and a deep (5 hidden layers of decreasing sizes) feedforward neural network $A$ (the more accurate and energy-costly model).
The confidence score $\alpha_G$ used for the cascading model $C$ corresponds to the fraction of training samples in the reached leaf node that belong to the predicted class.
We use $\varepsilon=0.2$ as a parameter for $C$, i.e., for an instance $\mathbf{x}$ we use the prediction of $G$ when the confidence score $\alpha_G(\mathbf{x})$ is at least $1-\varepsilon=0.8$, or the prediction of $A$ otherwise.
As an oracle for the routing model $R$ we use a logistic regressor with balanced class-weights (to account for potentially imbalanced classes coming from the predictions of $G$ and $A$ in the validation set).

For each of the competitors, we measure its \textit{accuracy} (i.e., fraction of correctly classified instances), \textit{total prediction time} (measured on the whole test set, in \unit{\milli\second}), and \textit{energy consumption} (measured on the whole test set, in \unit{\micro\watt\hour}).
Moreover, for our methods $C$ and $R$ we also report the fraction of predictions performed by the simpler component $G$ and the time overhead required for the model selection procedure itself\footnote{The energy overhead is not reported because it is minimal and difficult to measure accurately.}.
The running time is measured using the python function \texttt{time.perf\_counter()}.
The energy consumption is estimated using the python package \textit{CodeCarbon}\footnote{\url{https://codecarbon.io}}.
In order to smooth the time and energy measurements, they are averaged over 1\,000 predictions of the entire test set.
The example is run on a MacBook Air (M1, 16GB Memory) and the measurements are reported in Table~\ref{tab:classifier_comparison}.

\input{trunk/tab/table}

We observe that our hybrid methods $C$ and $R$, respectively based on cascading and routing, are effective in balancing accuracy and energy consumption relative to their components, the lightweight model $G$ and the accurate, energy-intensive model $A$. 
Importantly, the computational overhead introduced by the model selection process is negligible, adding only $\approx$0.2 \unit{\milli\second} to an average inference time of $\approx$30 \unit{\milli\second}.
Quantitatively, the cascading method $C$ retains 94.7\% of the accuracy of $A$ (a reduction of only 0.05), while improving inference speed by 21.25\% (7.95 \unit{\milli\second}) and reducing energy consumption by 21.27\% (8.68 \unit{\micro\watt\hour}). 
The routing method $R$ trades slightly more accuracy, retaining 89.7\% of $A$'s performance (a reduction of 0.1), for even greater energy efficiency, cutting consumption by 24.44\% (9.97 \unit{\micro\watt\hour}) and improving speed by 11.51\% (4.31 \unit{\milli\second}). 
These results highlight the potential of hybrid inference strategies to deliver substantial energy savings with only modest accuracy degradation.

\section{Discussion}
In this section, we discuss the key aspects we deem paramount to be considered while further developing the idea of Green AI model selection. More specifically, we cover aspects regarding the generalizability of our intuition, potential  impediments connected to porting the presented techniques to practice, and other nuances that may arise while further developing Green AI dynamic model selection.

\paragraph{On the generalizability to other tasks.}
While our study focuses on classification, the proposed approach can in principle be extended to other inference paradigms, including generative tasks, e.g., text generation and image synthesis, and large language models~\cite{bae2023fast,rahmath2024early}.
However, cascading in such settings poses new challenges: Unlike classification, generation tasks often produce variable-length outputs and lack a clear, standardized notion of ``confidence'', making it harder to decide when to accept early outputs. 
Furthermore, in autoregressive models, energy cost and quality are tightly coupled over long sequences, which complicates dynamic routing or early termination. 
Designing confidence surrogates or lightweight proxies for generation quality is an open and non-trivial research direction~\cite{xu2024benchmarking}.

\paragraph{On the consumption of the oracle.}
Routing strategies rely on a pre-trained oracle that predicts the most suitable model to use for a given input. 
While our results show that routing can outperform static selection in terms of energy efficiency, this advantage must be weighed against the energy and latency overhead introduced by the oracle itself. 
Although this cost can often be amortized, especially when the oracle is lightweight compared to the target models, it is nonetheless a relevant factor in real-world deployments and should be included in life-cycle assessments~\cite{zhang2018mixture}.

\paragraph{On the precision-energy consumption tradeoff.}
A central trade-off in energy-aware inference lies in balancing precision and energy consumption~\cite{brownlee2021exploring,kim2022tradeoff}. Tuning the confidence threshold $\epsilon$ in cascading or adjusting the routing oracle's decision boundary directly affects both the fraction of queries routed to low-cost models and the overall accuracy. 
This trade-off is highly application-dependent: For some critical tasks even minor drops in accuracy may be unacceptable, while for others tolerating occasional misclassifications may be worthwhile for substantial energy savings.

\paragraph{On the specificity of the task at hand.}
The benefits of dynamic model selection  depend heavily on the characteristics of the task. 
Tasks with highly skewed input difficulty, where many inputs are easily handled by simple models, stand to benefit most from cascading or routing~\cite{damani2024learning}. 
In contrast, tasks that are uniformly hard may offer little opportunity for savings, as most queries will require the most complex model regardless. 
This suggests that per-task calibration or meta-learning strategies could further enhance the adaptability of energy-aware approaches. Specific task might also dictate the dynamic model selection strategy. For example, by considering image generation quality and energy consumption~\cite{bertazzini2025hidden}, model routing might be to date the only solution~applicable.

\paragraph{On the energy cost of loading models.}
Energy measurement methodologies must carefully account for the cost of loading models into memory, especially when switching between models incurs overhead due to I/O or hardware constraints~\cite{JI2021102183}. 
In scenarios where models are not kept in memory persistently, the benefit of selecting a low-cost model may be offset by the loading cost. 
This points to the importance of deployment-aware design: in serverless or constrained edge environments, keeping a subset of models ``warm'' may be necessary for real energy savings.

\paragraph{On the development cost of maintaining models updated.}
The practical deployment of multi-model systems introduces non-trivial maintenance costs. 
Each model in the pool must be monitored, updated, and re-validated to cope with data drift, distribution shifts, or evolving application requirements. 
This adds complexity to the life-cycle management of the AI system and raises questions about the long-term cost–benefit balance. 
Approaches such as continual learning may help reduce redundancy and maintain performance with a smaller, more efficient model pool~\cite{majidi2024efficient}.

\paragraph{On the optimization of model carbon footprint.}
In this work we primarily focused on model energy consumption, i.e., the raw energy consumed by the models. In a broader perspective however, the carbon footprint of the models, i.e., the total amount of greenhouse gases required to produce the energy consumed by the models, might be instead the primary metric we want to optimize for~\cite{wang2025carbon}. 
Distinguishing between energy consumption and carbon footprint is necessary in contexts where models are not powered by the same energy grid, e.g., in a distributed deployment scenario. By considering the core intuition behind the presented methods, we argue that these can be effortlessly adapted by considering the measured carbon footprint of the models instead of the energy consumption considered in this contribution.

\section{Conclusions and Future Work}
With the great technological advancements AI brought, its evergrowing popularization, and its non-negligible environmental impact, we are responsible to conceive novel solutions that preserve technological advancements while optimizing environmental sustainability. In this work we present the concept of \textit{Green AI dynamic model selection}, which lives at the intersection of dynamic model selection and model environmental sustainability. The core contribution presented in this study is twofold, by proposing two distinct techniques through which Green AI can be achieved by dynamically selecting models according to their environmental sustainability. The two methods are referred to in this work as \textit{Green AI Dynamic Model Cascading} and \textit{Green AI Dynamic Model Routing}. To support our documented intuition, we report an empirical proof of concept, which showcases in practical terms the potential of the proposed idea. While the results of our experimentation are by no means to be considered as generalizable or conclusive, they support us in arguing that \textit{Green AI dynamic model selection} is a Green AI strategy that is worth to be further investigated. To further support our intuition, we also further delve into speculating on the core concepts, impediments, and benefits of \textit{Green AI dynamic model selection}, in the hope that our contribution can support other researchers in making AI greener.

%% file: trunk/tab/table.tex
\begin{table}[ht]
\centering
\caption{Performance and energy consumption of classifiers}
\begin{tabular}{lrrrrr}
\toprule
\textbf{Classifier} & \textbf{Fraction of $G$} & \textbf{Overhead (\unit{\milli\second})} & 
\textbf{Accuracy} & \textbf{Time (\unit{\milli\second})} & \textbf{Energy (\unit{\micro\watt\hour})} \\
\midrule
($G$) Decision Tree   & 1.00    &  0.00    &  0.73 & 0.13 & 0.13 \\
($A$) Neural Network  & 0.00    &  0.00    &  0.98 & 37.44 & 40.80 \\
\midrule
($C$) \textbf{Cascading}       & 0.65 &  0.22 &  0.92 & 29.48 & 32.12 \\
($R$) \textbf{Routing}         & 0.62 &  0.19 &  0.88 & 33.13 & 30.83 \\
\bottomrule
\end{tabular}
\label{tab:classifier_comparison}
\end{table}

%% file: main.bbl
\begin{thebibliography}{45}
\expandafter\ifx\csname natexlab\endcsname\relax\def\natexlab#1{#1}\fi
\providecommand{\url}[1]{\texttt{#1}}
\providecommand{\href}[2]{#2}
\providecommand{\path}[1]{#1}
\providecommand{\DOIprefix}{doi:}
\providecommand{\ArXivprefix}{arXiv:}
\providecommand{\URLprefix}{URL: }
\providecommand{\Pubmedprefix}{pmid:}
\providecommand{\doi}[1]{\href{http://dx.doi.org/#1}{\path{#1}}}
\providecommand{\Pubmed}[1]{\href{pmid:#1}{\path{#1}}}
\providecommand{\bibinfo}[2]{#2}
\ifx\xfnm\relax \def\xfnm[#1]{\unskip,\space#1}\fi
\bibitem[{Wu et~al.(2022)Wu, Raghavendra, Gupta, Acun, Ardalani, Maeng, Chang, Aga, Huang, Bai et~al.}]{wu2022sustainable}
\bibinfo{author}{C.-J. Wu}, \bibinfo{author}{R.~Raghavendra}, \bibinfo{author}{U.~Gupta}, \bibinfo{author}{B.~Acun}, \bibinfo{author}{N.~Ardalani}, \bibinfo{author}{K.~Maeng}, \bibinfo{author}{G.~Chang}, \bibinfo{author}{F.~Aga}, \bibinfo{author}{J.~Huang}, \bibinfo{author}{C.~Bai}, et~al.,
\newblock \bibinfo{title}{Sustainable ai: Environmental implications, challenges and opportunities},
\newblock \bibinfo{journal}{Proceedings of Machine Learning and Systems} \bibinfo{volume}{4} (\bibinfo{year}{2022}) \bibinfo{pages}{795--813}.
\bibitem[{Schwartz et~al.(2020)Schwartz, Dodge, Smith, and Etzioni}]{schwartz2020green}
\bibinfo{author}{R.~Schwartz}, \bibinfo{author}{J.~Dodge}, \bibinfo{author}{N.~A. Smith}, \bibinfo{author}{O.~Etzioni},
\newblock \bibinfo{title}{Green ai},
\newblock \bibinfo{journal}{Communications of the ACM} \bibinfo{volume}{63} (\bibinfo{year}{2020}) \bibinfo{pages}{54--63}.
\bibitem[{Verdecchia et~al.(2023)Verdecchia, Sallou, and Cruz}]{verdecchia2023systematic}
\bibinfo{author}{R.~Verdecchia}, \bibinfo{author}{J.~Sallou}, \bibinfo{author}{L.~Cruz},
\newblock \bibinfo{title}{A systematic review of green ai},
\newblock \bibinfo{journal}{Wiley Interdisciplinary Reviews: Data Mining and Knowledge Discovery} \bibinfo{volume}{13} (\bibinfo{year}{2023}) \bibinfo{pages}{e1507}.
\bibitem[{Cruz et~al.(2025)Cruz, Fernandes, Kirkeby, Mart{\'\i}nez-Fern{\'a}ndez, Sallou, Anwar, Barba~Roque, Bogner, Casta{\~n}o, Castor et~al.}]{cruz2025greening}
\bibinfo{author}{L.~Cruz}, \bibinfo{author}{J.~P. Fernandes}, \bibinfo{author}{M.~H. Kirkeby}, \bibinfo{author}{S.~Mart{\'\i}nez-Fern{\'a}ndez}, \bibinfo{author}{J.~Sallou}, \bibinfo{author}{H.~Anwar}, \bibinfo{author}{E.~Barba~Roque}, \bibinfo{author}{J.~Bogner}, \bibinfo{author}{J.~Casta{\~n}o}, \bibinfo{author}{F.~Castor}, et~al.,
\newblock \bibinfo{title}{Greening ai-enabled systems with software engineering: A research agenda for environmentally sustainable ai practices},
\newblock \bibinfo{journal}{ACM SIGSOFT Software Engineering Notes} \bibinfo{volume}{50} (\bibinfo{year}{2025}) \bibinfo{pages}{14--23}.
\bibitem[{Viola and Jones(2001)}]{viola2001cascade}
\bibinfo{author}{P.~Viola}, \bibinfo{author}{M.~Jones},
\newblock \bibinfo{title}{Rapid object detection using a boosted cascade of simple features},
\newblock in: \bibinfo{booktitle}{Proceedings of the 2001 IEEE Computer Society Conference on Computer Vision and Pattern Recognition. CVPR 2001}, volume~\bibinfo{volume}{1}, \bibinfo{year}{2001}, pp. \bibinfo{pages}{I--I}. \DOIprefix\doi{10.1109/CVPR.2001.990517}.
\bibitem[{Jacobs et~al.(1991)Jacobs, Jordan, Nowlan, and Hinton}]{jacobs1991moe}
\bibinfo{author}{R.~A. Jacobs}, \bibinfo{author}{M.~I. Jordan}, \bibinfo{author}{S.~J. Nowlan}, \bibinfo{author}{G.~E. Hinton},
\newblock \bibinfo{title}{Adaptive mixtures of local experts},
\newblock \bibinfo{journal}{Neural Computation} \bibinfo{volume}{3} (\bibinfo{year}{1991}) \bibinfo{pages}{79--87}. \DOIprefix\doi{10.1162/neco.1991.3.1.79}.
\bibitem[{Lacoste et~al.(2019)Lacoste, Luccioni, Schmidt, and Dandres}]{lacoste2019quantifying}
\bibinfo{author}{A.~Lacoste}, \bibinfo{author}{A.~Luccioni}, \bibinfo{author}{V.~Schmidt}, \bibinfo{author}{T.~Dandres},
\newblock \bibinfo{title}{Quantifying the carbon emissions of machine learning},
\newblock \bibinfo{journal}{arXiv preprint arXiv:1910.09700}  (\bibinfo{year}{2019}).
\bibitem[{Strubell et~al.(2020)Strubell, Ganesh, and McCallum}]{strubell2020energy}
\bibinfo{author}{E.~Strubell}, \bibinfo{author}{A.~Ganesh}, \bibinfo{author}{A.~McCallum},
\newblock \bibinfo{title}{Energy and policy considerations for modern deep learning research},
\newblock in: \bibinfo{booktitle}{Proceedings of the AAAI conference on artificial intelligence}, volume~\bibinfo{volume}{34}, \bibinfo{year}{2020}, pp. \bibinfo{pages}{13693--13696}.
\bibitem[{Rouhani et~al.(2016)Rouhani, Mirhoseini, and Koushanfar}]{rouhani2016delight}
\bibinfo{author}{B.~D. Rouhani}, \bibinfo{author}{A.~Mirhoseini}, \bibinfo{author}{F.~Koushanfar},
\newblock \bibinfo{title}{Delight: Adding energy dimension to deep neural networks},
\newblock in: \bibinfo{booktitle}{Proceedings of the 2016 International Symposium on Low Power Electronics and Design}, \bibinfo{year}{2016}, pp. \bibinfo{pages}{112--117}.
\bibitem[{Garc{\'\i}a-Mart{\'\i}n et~al.(2021)Garc{\'\i}a-Mart{\'\i}n, Lavesson, Grahn, Casalicchio, and Boeva}]{garcia2021energy}
\bibinfo{author}{E.~Garc{\'\i}a-Mart{\'\i}n}, \bibinfo{author}{N.~Lavesson}, \bibinfo{author}{H.~Grahn}, \bibinfo{author}{E.~Casalicchio}, \bibinfo{author}{V.~Boeva},
\newblock \bibinfo{title}{Energy-aware very fast decision tree},
\newblock \bibinfo{journal}{International Journal of Data Science and Analytics} \bibinfo{volume}{11} (\bibinfo{year}{2021}) \bibinfo{pages}{105--126}.
\bibitem[{Rungsuptaweekoon et~al.(2017)Rungsuptaweekoon, Visoottiviseth, and Takano}]{rungsuptaweekoon2017evaluating}
\bibinfo{author}{K.~Rungsuptaweekoon}, \bibinfo{author}{V.~Visoottiviseth}, \bibinfo{author}{R.~Takano},
\newblock \bibinfo{title}{Evaluating the power efficiency of deep learning inference on embedded gpu systems},
\newblock in: \bibinfo{booktitle}{2017 2nd international conference on information technology (INCIT)}, \bibinfo{organization}{IEEE}, \bibinfo{year}{2017}, pp. \bibinfo{pages}{1--5}.
\bibitem[{Yang et~al.(2018)Yang, Zhu, and Liu}]{yang2018energy}
\bibinfo{author}{H.~Yang}, \bibinfo{author}{Y.~Zhu}, \bibinfo{author}{J.~Liu},
\newblock \bibinfo{title}{Energy-constrained compression for deep neural networks via weighted sparse projection and layer input masking},
\newblock \bibinfo{journal}{arXiv preprint arXiv:1806.04321}  (\bibinfo{year}{2018}).
\bibitem[{Marini et~al.(2025)Marini, Pampaloni, Di~Martino, Verdecchia, and Vicario}]{11039308}
\bibinfo{author}{N.~Marini}, \bibinfo{author}{L.~Pampaloni}, \bibinfo{author}{F.~Di~Martino}, \bibinfo{author}{R.~Verdecchia}, \bibinfo{author}{E.~Vicario},
\newblock \bibinfo{title}{{ Green AI: Which Programming Language Consumes the Most? }},
\newblock in: \bibinfo{booktitle}{2025 IEEE/ACM 9th International Workshop on Green and Sustainable Software (GREENS)}, \bibinfo{publisher}{IEEE Computer Society}, \bibinfo{address}{Los Alamitos, CA, USA}, \bibinfo{year}{2025}, pp. \bibinfo{pages}{12--19}. \URLprefix \url{https://doi.ieeecomputersociety.org/10.1109/GREENS66463.2025.00007}. \DOIprefix\doi{10.1109/GREENS66463.2025.00007}.
\bibitem[{Georgiou et~al.(2022)Georgiou, Kechagia, Sharma, Sarro, and Zou}]{georgiou2022green}
\bibinfo{author}{S.~Georgiou}, \bibinfo{author}{M.~Kechagia}, \bibinfo{author}{T.~Sharma}, \bibinfo{author}{F.~Sarro}, \bibinfo{author}{Y.~Zou},
\newblock \bibinfo{title}{Green ai: Do deep learning frameworks have different costs?},
\newblock in: \bibinfo{booktitle}{Proceedings of the 44th International Conference on Software Engineering}, \bibinfo{year}{2022}, pp. \bibinfo{pages}{1082--1094}.
\bibitem[{de~Chavannes et~al.(2021)de~Chavannes, Kongsbak, Rantzau, and Derczynski}]{de2021hyperparameter}
\bibinfo{author}{L.~H.~P. de~Chavannes}, \bibinfo{author}{M.~G.~K. Kongsbak}, \bibinfo{author}{T.~Rantzau}, \bibinfo{author}{L.~Derczynski},
\newblock \bibinfo{title}{Hyperparameter power impact in transformer language model training},
\newblock in: \bibinfo{booktitle}{Proceedings of the second workshop on simple and efficient natural language processing}, \bibinfo{year}{2021}, pp. \bibinfo{pages}{96--118}.
\bibitem[{Magno et~al.(2017)Magno, Pritz, Mayer, and Benini}]{magno2017deepemote}
\bibinfo{author}{M.~Magno}, \bibinfo{author}{M.~Pritz}, \bibinfo{author}{P.~Mayer}, \bibinfo{author}{L.~Benini},
\newblock \bibinfo{title}{Deepemote: Towards multi-layer neural networks in a low power wearable multi-sensors bracelet},
\newblock in: \bibinfo{booktitle}{2017 7th IEEE international workshop on advances in sensors and interfaces (IWASI)}, \bibinfo{organization}{IEEE}, \bibinfo{year}{2017}, pp. \bibinfo{pages}{32--37}.
\bibitem[{Barlaud and Guyard(2021)}]{barlaud2021learning}
\bibinfo{author}{M.~Barlaud}, \bibinfo{author}{F.~Guyard},
\newblock \bibinfo{title}{Learning sparse deep neural networks using efficient structured projections on convex constraints for green ai},
\newblock in: \bibinfo{booktitle}{2020 25th international conference on pattern recognition (ICPR)}, \bibinfo{organization}{IEEE}, \bibinfo{year}{2021}, pp. \bibinfo{pages}{1566--1573}.
\bibitem[{Stamoulis et~al.(2018)Stamoulis, Chin, Prakash, Fang, Sajja, Bognar, and Marculescu}]{stamoulis2018designing}
\bibinfo{author}{D.~Stamoulis}, \bibinfo{author}{T.-W.~R. Chin}, \bibinfo{author}{A.~K. Prakash}, \bibinfo{author}{H.~Fang}, \bibinfo{author}{S.~Sajja}, \bibinfo{author}{M.~Bognar}, \bibinfo{author}{D.~Marculescu},
\newblock \bibinfo{title}{Designing adaptive neural networks for energy-constrained image classification},
\newblock in: \bibinfo{booktitle}{2018 IEEE/ACM International Conference on Computer-Aided Design (ICCAD)}, \bibinfo{organization}{IEEE}, \bibinfo{year}{2018}, pp. \bibinfo{pages}{1--8}.
\bibitem[{Sponner et~al.(2024)Sponner, Waschneck, and Kumar}]{sponner2024adapting}
\bibinfo{author}{M.~Sponner}, \bibinfo{author}{B.~Waschneck}, \bibinfo{author}{A.~Kumar},
\newblock \bibinfo{title}{Adapting neural networks at runtime: Current trends in at-runtime optimizations for deep learning},
\newblock \bibinfo{journal}{ACM Computing Surveys} \bibinfo{volume}{56} (\bibinfo{year}{2024}) \bibinfo{pages}{1--40}.
\bibitem[{Gondi and Pratap(2021)}]{gondi2021performance}
\bibinfo{author}{S.~Gondi}, \bibinfo{author}{V.~Pratap},
\newblock \bibinfo{title}{Performance and efficiency evaluation of asr inference on the edge},
\newblock \bibinfo{journal}{Sustainability} \bibinfo{volume}{13} (\bibinfo{year}{2021}) \bibinfo{pages}{12392}.
\bibitem[{Yang et~al.(2020)Yang, Hua, Shi, Wang, Zhang, and Letaief}]{yang2020sparse}
\bibinfo{author}{X.~Yang}, \bibinfo{author}{S.~Hua}, \bibinfo{author}{Y.~Shi}, \bibinfo{author}{H.~Wang}, \bibinfo{author}{J.~Zhang}, \bibinfo{author}{K.~B. Letaief},
\newblock \bibinfo{title}{Sparse optimization for green edge ai inference},
\newblock \bibinfo{journal}{Journal of communications and information networks} \bibinfo{volume}{5} (\bibinfo{year}{2020}) \bibinfo{pages}{1--15}.
\bibitem[{Yosuf et~al.(2021)Yosuf, Mohamed, Alenazi, El-Gorashi, and Elmirghani}]{yosuf2021energy}
\bibinfo{author}{B.~A. Yosuf}, \bibinfo{author}{S.~H. Mohamed}, \bibinfo{author}{M.~M. Alenazi}, \bibinfo{author}{T.~E. El-Gorashi}, \bibinfo{author}{J.~M. Elmirghani},
\newblock \bibinfo{title}{Energy-efficient ai over a virtualized cloud fog network},
\newblock in: \bibinfo{booktitle}{Proceedings of the twelfth ACM international conference on future energy systems}, \bibinfo{year}{2021}, pp. \bibinfo{pages}{328--334}.
\bibitem[{G{\"u}ler and Yener(2021)}]{guler2021energy}
\bibinfo{author}{B.~G{\"u}ler}, \bibinfo{author}{A.~Yener},
\newblock \bibinfo{title}{Energy-harvesting distributed machine learning},
\newblock in: \bibinfo{booktitle}{2021 IEEE international symposium on information theory (ISIT)}, \bibinfo{organization}{IEEE}, \bibinfo{year}{2021}, pp. \bibinfo{pages}{320--325}.
\bibitem[{Verdecchia et~al.(2022)Verdecchia, Cruz, Sallou, Lin, Wickenden, and Hotellier}]{verdecchia2022data}
\bibinfo{author}{R.~Verdecchia}, \bibinfo{author}{L.~Cruz}, \bibinfo{author}{J.~Sallou}, \bibinfo{author}{M.~Lin}, \bibinfo{author}{J.~Wickenden}, \bibinfo{author}{E.~Hotellier},
\newblock \bibinfo{title}{Data-centric green ai an exploratory empirical study},
\newblock in: \bibinfo{booktitle}{2022 international conference on ICT for sustainability (ICT4S)}, \bibinfo{organization}{IEEE}, \bibinfo{year}{2022}, pp. \bibinfo{pages}{35--45}.
\bibitem[{Salehi and Schmeink(2023)}]{salehi2023data}
\bibinfo{author}{S.~Salehi}, \bibinfo{author}{A.~Schmeink},
\newblock \bibinfo{title}{Data-centric green artificial intelligence: A survey},
\newblock \bibinfo{journal}{IEEE Transactions on Artificial Intelligence} \bibinfo{volume}{5} (\bibinfo{year}{2023}) \bibinfo{pages}{1973--1989}.
\bibitem[{Alswaitti et~al.(2025)Alswaitti, Verdecchia, Danoy, Bouvry, and Pecero}]{alswaitti2025training}
\bibinfo{author}{M.~Alswaitti}, \bibinfo{author}{R.~Verdecchia}, \bibinfo{author}{G.~Danoy}, \bibinfo{author}{P.~Bouvry}, \bibinfo{author}{J.~Pecero},
\newblock \bibinfo{title}{Training green ai models using elite samples},
\newblock \bibinfo{journal}{IEEE Transactions on Sustainable Computing}  (\bibinfo{year}{2025}).
\bibitem[{Kumar et~al.(2024)Kumar, Datta, Singh, Singh, and Sharma}]{kumar2024opportunities}
\bibinfo{author}{S.~Kumar}, \bibinfo{author}{S.~Datta}, \bibinfo{author}{V.~Singh}, \bibinfo{author}{S.~K. Singh}, \bibinfo{author}{R.~Sharma},
\newblock \bibinfo{title}{Opportunities and challenges in data-centric ai},
\newblock \bibinfo{journal}{IEEE Access}  (\bibinfo{year}{2024}).
\bibitem[{J{\"a}rvenp{\"a}{\"a} et~al.(2024)J{\"a}rvenp{\"a}{\"a}, Lago, Bogner, Lewis, Muccini, and Ozkaya}]{jarvenpaa2024synthesis}
\bibinfo{author}{H.~J{\"a}rvenp{\"a}{\"a}}, \bibinfo{author}{P.~Lago}, \bibinfo{author}{J.~Bogner}, \bibinfo{author}{G.~Lewis}, \bibinfo{author}{H.~Muccini}, \bibinfo{author}{I.~Ozkaya},
\newblock \bibinfo{title}{A synthesis of green architectural tactics for ml-enabled systems},
\newblock in: \bibinfo{booktitle}{Proceedings of the 46th International Conference on Software Engineering: Software Engineering in Society}, \bibinfo{year}{2024}, pp. \bibinfo{pages}{130--141}.
\bibitem[{Nijkamp et~al.(2024)Nijkamp, Sallou, van~der Heijden, and Cruz}]{nijkamp2024green}
\bibinfo{author}{N.~Nijkamp}, \bibinfo{author}{J.~Sallou}, \bibinfo{author}{N.~van~der Heijden}, \bibinfo{author}{L.~Cruz},
\newblock \bibinfo{title}{Green ai in action: Strategic model selection for ensembles in production},
\newblock in: \bibinfo{booktitle}{Proceedings of the 1st ACM International Conference on AI-Powered Software}, \bibinfo{year}{2024}, pp. \bibinfo{pages}{50--58}.
\bibitem[{Omar et~al.(2024)Omar, Bogner, Muccini, Lago, Mart{\'\i}nez-Fern{\'a}ndez, and Franch}]{omar2024more}
\bibinfo{author}{R.~Omar}, \bibinfo{author}{J.~Bogner}, \bibinfo{author}{H.~Muccini}, \bibinfo{author}{P.~Lago}, \bibinfo{author}{S.~Mart{\'\i}nez-Fern{\'a}ndez}, \bibinfo{author}{X.~Franch},
\newblock \bibinfo{title}{The more the merrier? navigating accuracy vs. energy efficiency design trade-offs in ensemble learning systems},
\newblock \bibinfo{journal}{arXiv preprint arXiv:2407.02914}  (\bibinfo{year}{2024}).
\bibitem[{Matathammal et~al.(2025)Matathammal, Gupta, Lavanya, Halgatti, Gupta, and Vaidhyanathan}]{matathammal2025edgemlbalancer}
\bibinfo{author}{A.~Matathammal}, \bibinfo{author}{K.~Gupta}, \bibinfo{author}{L.~Lavanya}, \bibinfo{author}{A.~V. Halgatti}, \bibinfo{author}{P.~Gupta}, \bibinfo{author}{K.~Vaidhyanathan},
\newblock \bibinfo{title}{Edgemlbalancer: A self-adaptive approach for dynamic model switching on resource-constrained edge devices},
\newblock \bibinfo{journal}{arXiv preprint arXiv:2502.06493}  (\bibinfo{year}{2025}).
\bibitem[{Anderson and Burnham(2004)}]{anderson2004model}
\bibinfo{author}{D.~Anderson}, \bibinfo{author}{K.~Burnham},
\newblock \bibinfo{title}{Model selection and multi-model inference},
\newblock \bibinfo{journal}{Second. NY: Springer-Verlag} \bibinfo{volume}{63} (\bibinfo{year}{2004}) \bibinfo{pages}{10}.
\bibitem[{Merz(1996)}]{merz1996dynamical}
\bibinfo{author}{C.~J. Merz},
\newblock \bibinfo{title}{Dynamical selection of learning algorithms},
\newblock in: \bibinfo{booktitle}{Learning from Data: Artificial Intelligence and Statistics V}, \bibinfo{publisher}{Springer}, \bibinfo{year}{1996}, pp. \bibinfo{pages}{281--290}.
\bibitem[{Armstrong et~al.(2006)Armstrong, Christen, McCreath, and Rendell}]{armstrong2006dynamic}
\bibinfo{author}{W.~Armstrong}, \bibinfo{author}{P.~Christen}, \bibinfo{author}{E.~McCreath}, \bibinfo{author}{A.~P. Rendell},
\newblock \bibinfo{title}{Dynamic algorithm selection using reinforcement learning},
\newblock in: \bibinfo{booktitle}{2006 international workshop on integrating ai and data mining}, \bibinfo{organization}{IEEE}, \bibinfo{year}{2006}, pp. \bibinfo{pages}{18--25}.
\bibitem[{Bae et~al.(2023)Bae, Ko, Song, and Yun}]{bae2023fast}
\bibinfo{author}{S.~Bae}, \bibinfo{author}{J.~Ko}, \bibinfo{author}{H.~Song}, \bibinfo{author}{S.-Y. Yun},
\newblock \bibinfo{title}{Fast and robust early-exiting framework for autoregressive language models with synchronized parallel decoding},
\newblock \bibinfo{journal}{arXiv preprint arXiv:2310.05424}  (\bibinfo{year}{2023}).
\bibitem[{Rahmath~P et~al.(2024)Rahmath~P, Srivastava, Chaurasia, Pacheco, and Couto}]{rahmath2024early}
\bibinfo{author}{H.~Rahmath~P}, \bibinfo{author}{V.~Srivastava}, \bibinfo{author}{K.~Chaurasia}, \bibinfo{author}{R.~G. Pacheco}, \bibinfo{author}{R.~S. Couto},
\newblock \bibinfo{title}{Early-exit deep neural network-a comprehensive survey},
\newblock \bibinfo{journal}{ACM Computing Surveys} \bibinfo{volume}{57} (\bibinfo{year}{2024}) \bibinfo{pages}{1--37}.
\bibitem[{Xu et~al.(2024)Xu, Lu, Schoenebeck, and Kong}]{xu2024benchmarking}
\bibinfo{author}{S.~Xu}, \bibinfo{author}{Y.~Lu}, \bibinfo{author}{G.~Schoenebeck}, \bibinfo{author}{Y.~Kong},
\newblock \bibinfo{title}{Benchmarking llms' judgments with no gold standard},
\newblock \bibinfo{journal}{arXiv preprint arXiv:2411.07127}  (\bibinfo{year}{2024}).
\bibitem[{Zhang et~al.(2018)Zhang, Davoodi, and Hu}]{zhang2018mixture}
\bibinfo{author}{B.~Zhang}, \bibinfo{author}{A.~Davoodi}, \bibinfo{author}{Y.-H. Hu},
\newblock \bibinfo{title}{A mixture of expert approach for low-cost customization of deep neural networks},
\newblock \bibinfo{journal}{arXiv preprint arXiv:1811.00056}  (\bibinfo{year}{2018}).
\bibitem[{Brownlee et~al.(2021)Brownlee, Adair, Haraldsson, and Jabbo}]{brownlee2021exploring}
\bibinfo{author}{A.~E. Brownlee}, \bibinfo{author}{J.~Adair}, \bibinfo{author}{S.~O. Haraldsson}, \bibinfo{author}{J.~Jabbo},
\newblock \bibinfo{title}{Exploring the accuracy--energy trade-off in machine learning},
\newblock in: \bibinfo{booktitle}{2021 IEEE/ACM International Workshop on Genetic Improvement (GI)}, \bibinfo{organization}{IEEE}, \bibinfo{year}{2021}, pp. \bibinfo{pages}{11--18}.
\bibitem[{Kim et~al.(2022)Kim, Saad, Mozaffari, and Debbah}]{kim2022tradeoff}
\bibinfo{author}{M.~Kim}, \bibinfo{author}{W.~Saad}, \bibinfo{author}{M.~Mozaffari}, \bibinfo{author}{M.~Debbah},
\newblock \bibinfo{title}{On the tradeoff between energy, precision, and accuracy in federated quantized neural networks},
\newblock in: \bibinfo{booktitle}{ICC 2022-IEEE International Conference on Communications}, \bibinfo{organization}{IEEE}, \bibinfo{year}{2022}, pp. \bibinfo{pages}{2194--2199}.
\bibitem[{Damani et~al.(2024)Damani, Shenfeld, Peng, Bobu, and Andreas}]{damani2024learning}
\bibinfo{author}{M.~Damani}, \bibinfo{author}{I.~Shenfeld}, \bibinfo{author}{A.~Peng}, \bibinfo{author}{A.~Bobu}, \bibinfo{author}{J.~Andreas},
\newblock \bibinfo{title}{Learning how hard to think: Input-adaptive allocation of lm computation},
\newblock \bibinfo{journal}{arXiv preprint arXiv:2410.04707}  (\bibinfo{year}{2024}).
\bibitem[{Bertazzini et~al.(2025)Bertazzini, Albisani, Baracchi, Shullani, and Verdecchia}]{bertazzini2025hidden}
\bibinfo{author}{G.~Bertazzini}, \bibinfo{author}{C.~Albisani}, \bibinfo{author}{D.~Baracchi}, \bibinfo{author}{D.~Shullani}, \bibinfo{author}{R.~Verdecchia},
\newblock \bibinfo{title}{{The Hidden Cost of an Image: Quantifying the Energy Consumption of AI Image Generation}},
\newblock \bibinfo{journal}{arXiv preprint arXiv:2506.17016}  (\bibinfo{year}{2025}).
\bibitem[{Ji et~al.(2021)Ji, Wu, Zhu, Chang, Liu, and Zhai}]{JI2021102183}
\bibinfo{author}{C.~Ji}, \bibinfo{author}{F.~Wu}, \bibinfo{author}{Z.~Zhu}, \bibinfo{author}{L.-P. Chang}, \bibinfo{author}{H.~Liu}, \bibinfo{author}{W.~Zhai},
\newblock \bibinfo{title}{Memory-efficient deep learning inference with incremental weight loading and data layout reorganization on edge systems},
\newblock \bibinfo{journal}{Journal of Systems Architecture} \bibinfo{volume}{118} (\bibinfo{year}{2021}) \bibinfo{pages}{102183}.
\bibitem[{Majidi et~al.(2024)Majidi, Khomh, Li, and Nikanjam}]{majidi2024efficient}
\bibinfo{author}{F.~Majidi}, \bibinfo{author}{F.~Khomh}, \bibinfo{author}{H.~Li}, \bibinfo{author}{A.~Nikanjam},
\newblock \bibinfo{title}{An efficient model maintenance approach for mlops},
\newblock \bibinfo{journal}{arXiv preprint arXiv:2412.04657}  (\bibinfo{year}{2024}).
\bibitem[{Wang et~al.(2025)Wang, Ardalani, Elhoushi, Jiang, Hsia, Sumbul, Mahajan, Wu, and Acun}]{wang2025carbon}
\bibinfo{author}{I.~Wang}, \bibinfo{author}{N.~Ardalani}, \bibinfo{author}{M.~Elhoushi}, \bibinfo{author}{D.~Jiang}, \bibinfo{author}{S.~Hsia}, \bibinfo{author}{E.~Sumbul}, \bibinfo{author}{D.~Mahajan}, \bibinfo{author}{C.-J. Wu}, \bibinfo{author}{B.~Acun},
\newblock \bibinfo{title}{Carbon aware transformers through joint model-hardware optimization},
\newblock \bibinfo{journal}{arXiv preprint arXiv:2505.01386}  (\bibinfo{year}{2025}).

\end{thebibliography}
